\documentclass[twocolumn,superscriptaddress]{revtex4}
\usepackage{graphicx}
\usepackage{color}

\begin{document}
\title{All-optical modulation in wavelength-sized epsilon-near-zero media}

\author{Alessandro Ciattoni}
\affiliation{Consiglio Nazionale delle Ricerche, CNR-SPIN, Via Vetoio 10, 67100 L'Aquila, Italy}

\author{Andrea Marini}
\affiliation{ICFO-Institut de Ciencies Fotoniques, The Barcelona Institute of Science and Technology, 08860 Castelldefels (Barcelona), Spain}

\author{Carlo Rizza}
\affiliation{Dipartimento di Scienza e Alta Tecnologia, Universit\`a degli studi dell'Insubria, Via Valleggio 11, 22100 Como, Italy}
\affiliation{Consiglio Nazionale delle Ricerche, CNR-SPIN, Via Vetoio 10, 67100 L'Aquila, Italy}

\begin{abstract}
We investigate the interaction of two pulses (pump and probe) scattered by a nonlinear epsilon-near-zero (ENZ) slab whose thickness is comparable with the ENZ wavelength. We show that when the probe has a narrow spectrum localized around the ENZ wavelength its transmission is dramatically affected by the intensity of the pump. Conversely, if the probe is not in the ENZ regime, its propagation is not noticeably affected by the pump. Such all-optical modulation is due to the oversensitive character of the ENZ regime and it is so efficient to even occur in a wavelength thick slab.
\end{abstract}

\maketitle


Materials exhibiting very small dielectric permittivity, or epsilon-near-zero (ENZ) media, have attracted a large research interest in the last decade \cite{Mahoud} since they host a regime where the electromagnetic field is spatially slowly-varying over a physically large region and thus amenable to be manipulated down to its finest details. This is a very basic and general mechanism which can be put at work in a number of different setups thus justifying the existing plethora of different effects and applications based on the ENZ regime. Relevant examples of ENZ phenomena are squeezing of the field through ultra-narrow channels \cite{Silve1}, tailoring of the radiation phase pattern of antennas \cite{Aluuu1}, unusual dynamics of surface and bulk polaritonic modes \cite{Ciatt3,Newman}, perfect absorption \cite{Zhongg,Yoonnn} enhancement of nonlocal effects \cite{Pollar,Davidd,Rizza1}, strong Purcell effect \cite{Chebyk} and unusual Goos-H\"{a}nchen effect \cite{Xuuuuu}. The smallness of the dielectric permittivity allows matter nonlinearity to effectively trigger a dielectric-metallic transition of the medium response \cite{Husako}, a mechanism supporting novel families of solitons \cite{Ciatt4} with exotic features like transverse power flow reversing \cite{Ciatt5} and frozen light \cite{Marini}. In addition, ENZ media provide the non-resonant enhancement of the normal electric field component across the vacuum-ENZ medium interface \cite{Campi2} and this yields relevant effects as transmissivity directional hysteresis \cite{Ciatt6} and enhancement of second and third harmonic generation  \cite{Vince1,Ciatt8}. Narrow ENZ plasmonic channels have been shown to host a different field enhancement mechanism \cite{Argyr1} producing marked nonlinear optical effect as temporal soliton excitation \cite{Argyr2}, and enhancement of second-harmonic generation efficiency \cite{Argyr3}. The intensified effect of matter nonlinearity in the ENZ regime, stemming from the spatially slowly-varying character of a pulse in such regime, has recently lead to the prediction of a marked self-interaction of pulses occurring in thin ENZ slabs \cite{Ciatt9}. Even though the most of the above intriguing ENZ mechanisms and effects have been predicted and observed in metamaterials, an increasing  research interest has been recently focused on plasmonic materials having a zero-crossing point of the permittivity real part close to their plasma frequency. Examples are semiconductors \cite{Junnnn}, strontium ruthenate \cite{Braicc}, aluminium doped zinc oxide \cite{Kinsey} and, probably the most investigated one, indium tin oxide \cite{Parkkk,Lukkkk,Capret,Kimmmm,Travis,Zhaooo}. Such plasmonic materials are intrinsically tunable since their plasma frequency can be varied using electrical or optical methods, and hence the ENZ frequency and bandwidth can be suitably adjusted for designing novel plasmonic devices with optical steering functionality.

In this Letter, we consider the nonlinear interaction of two pulses occurring in a very thin ENZ slab. The two quasi-monochromatic pulses, pump and probe with well separated spectral profiles, have very different powers and are launched simultaneously into the slab. Our full-wave simulations show that the transmission of the weak probe is affected by the pump only if it is spectrally located at the ENZ frequency. Such a marked all-optical pulse modulation is due to the fact that the pump produces an effective intensity dependent nonlinear shift of ENZ point so that the probe in the ENZ regime experiences slab dielectric-like and metallic-like behaviors at different pump intensities.

We consider the scattering setup sketched in Fig.1(a), where two co-propagating electromagnetic pulses are launched from vacuum to orthogonally impinge on the surface of a dielectric slab. We assume the dynamics of the slab polarization $\textbf{P}$ to be described by the equation \cite{Kogaaa,Contii}
\begin{equation} \label{polariz}
\frac{\partial^2 \textbf{P}}{\partial t^2}
        + \delta_e \omega_e \frac{\partial \textbf{P}}{\partial t}
        + \omega_e^2 \left( 1 + \frac{ \left| \textbf{P} \right|^2}{P_s^2} \right)^{-3/2} \textbf{P} = \epsilon_0 \left( \epsilon_s - 1 \right) \omega_e^2 \textbf{E}
\end{equation}
where $P_s$ is the saturation polarization that governs nonlinear oscillator behavior and $\textbf{E}$ is the radiation field. This model is justified by the fact that, for $|\textbf{P}|$ much smaller than $P_s$, Eq.(\ref{polariz}) reproduces the standard Kerr anharmonic equation and for larger $|\textbf{P}|$ it accounts for physically important higher order nonlinear terms (e.g. quintic contributions and saturation) \cite{Janyan}. Up to the zeroth order in $|\textbf{P}|/P_s$, Eq.(\ref{polariz}) reproduces the single-pole Lorentz oscillator with resonant frequency $\omega_e$, loss coefficient $\delta_e \omega_e$ and static dielectric permittivity $\epsilon_s$. Therefore, in the linear regime, the dielectric permittivity experienced by monochromatic $\exp \left(-i \omega t\right)$ fields is
\begin{equation} \label{eps}
\epsilon(\omega) = 1 + \frac{\epsilon_s -1}{1 - i \delta_e \left( \frac{\omega}{\omega_e} \right) - \left( \frac{\omega}{\omega_e} \right)^2},
\end{equation}
\begin{figure}
\center
\includegraphics[width=0.47\textwidth]{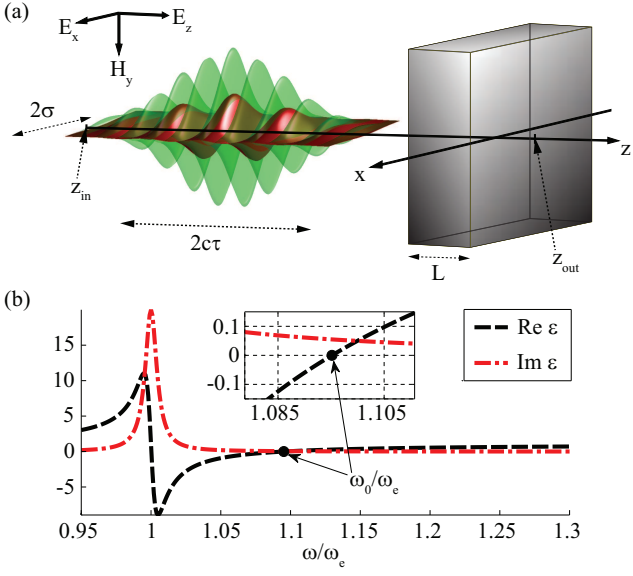}
\caption{(Color online) (a) Geometry of the scattering process. (b) Real and imaginary parts of the slab linear permittivity $\epsilon (\omega)$ of Eq.(\ref{eps}), zero-crossing-point $\textrm{Re} \left[ \epsilon \left( \omega_0 \right) \right] = 0$ at the frequency $\omega_0$ and ENZ regime $|\epsilon| \ll 1$ around $\omega_0$ (inset).}
\end{figure}
and it admits the zero-crossing-point $\textrm{Re} \left[ \epsilon \left( \omega_0 \right) \right] = 0$ at the frequency
\begin{equation} \label{omega0}
\omega_0 =  \frac{\omega_e}{\sqrt{2}} \left\{  \left(\epsilon_s +1 - \delta_e^2 \right) +  \left[ \left( \epsilon_s +1 - \delta_e^2 \right)^2 - 4 \epsilon_s \right]^{1/2}  \right\}^{1/2}.
\end{equation}
We have set $L = 1.25 \lambda_e$ for the slab thickness ($\lambda_e =  2 \pi c / \omega_e $ is the resonant wavelength) and $\delta_e = 0.01$, $\epsilon_s = 1.2$ for its dispersion parameters. As a consequence the slab supports the ENZ regime $|\epsilon| \ll 1$ around $\omega_0 = 1.095 \omega_e$ since, as reported in Fig.1(b) and in its inset, the imaginary part of the permittivity is small for frequencies close to $\omega_0$. The pump and probe are transverse magnetic (TM) pulses whose transverse electric field profile at the launching plane is $E_x(x,z_{in},t) = \left[ E_{pum} \sin \left( \omega_{pum} t \right) + E_{pro} \sin \left( \omega_{pro} t \right) \right] \exp \left[ -\frac{x^2}{\sigma^2} -\frac{(t-t_0)^2}{\tau^2} \right]$, where $t_0 = 3.178 \cdot 10^3 \omega_e^{-1}$ is a time shift. The pulses are both spatially and temporally localized, with spatial and temporal half-widths $\sigma = 1.25 \lambda_e$ and $\tau = 1.059 \cdot 10^3 \omega_e^{-1}$. Besides they are quasi-monochromatic since the carrier frequencies $\omega_{pum}$ and $\omega_{pro}$ will be chosen, in the simulations below, to be comparable with $\omega_e$ and the spectral width $\delta \omega \simeq 1 / (2\tau) = 4.721 \cdot 10^{-4} \omega_e$ of the pulses is much smaller than $\omega_e$. The scattering of the pulses by the slab was simulated by solving Maxwell equations coupled to Eq.(\ref{polariz}) using a home-made finite-difference time-domain code suitable for dealing with transverse magnetic pulses.

\begin{figure*}
\center
\includegraphics[width=1\textwidth]{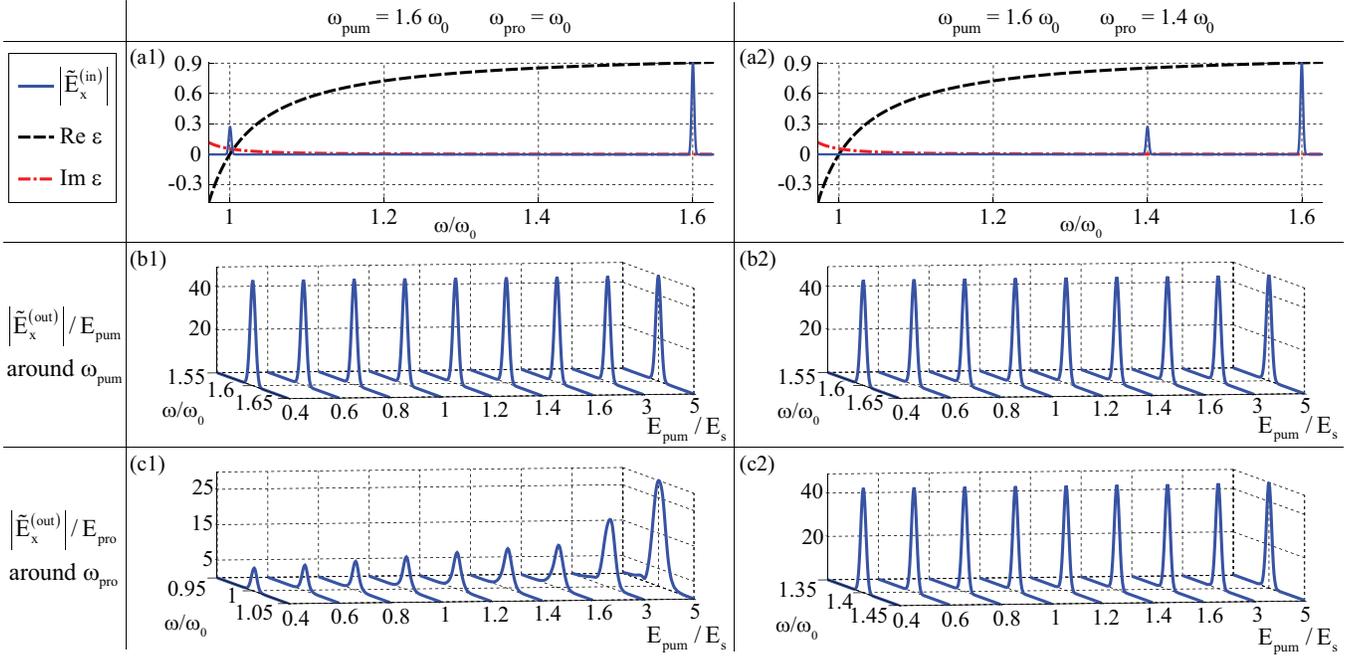}
\caption{(Color online) Left and right columns refer to the situations where the probe is in and out the ENZ regime, respectively. (a1) and (a2) spectra (arbitrary units) of the launched pump and probe at the plane $z_{in}$ superimposed to real and imaginary parts of the slab dielectric permittivity. (b1) and (b2) transmitted pump spectra normalized to the pump amplitude. (c1) and (c2) transmitted probe spectra normalized to the probe amplitude.}
\end{figure*}

In the first set of simulations we have set $\omega_{pum} = 1.6 \omega_0$, $E_{pum} /E_s = \left\{ 0.4, 0.6, 0.8, 1, 1.2, 1.4, 1.6, 3, 5 \right\}$, $\omega_{pro} = \omega_0$, $E_{pro} / E_s = 1 \cdot 10^{-3}$ where $E_s = P_s / \epsilon_0$. In this situation, the strong pump pulses (with different amplitudes) are out of the ENZ regime whereas the weak probe pulses (with fixed amplitude) is in the ENZ regime. The relevant results of the simulations are reported in the left column of Fig.2. In Fig.2(a1) we have plotted the absolute value of the Fourier transform $\tilde{E}_x^{(in)}$ (arbitrary units) of the incoming electric field $E_x (0,z_{in},t)$ superimposed to the real and imaginary parts of the permittivity and it is evident that two pulses are distinct excitations since their spectral profiles are well separated. In Fig.2(b1) we have plotted the absolute value of the Fourier transform $\tilde{E}_x^{(out)}$ of the outcoming electric field $E_x (0,z_{out},t)$ normalized with the pump amplitude $E_{pum}$, for frequencies around the pump frequency $\omega_{pum}$, and we note that the spectrum of the transmitted pump actually does not depend on its amplitude $E_{pum}$. In Fig.2(c1) we have plotted $|\tilde{E}_x^{(out)}|$ normalized with the probe amplitude $E_{pro}$, for frequencies around the probe frequency $\omega_{pro}$. The remarkable dependence of the spectrum of the transmitted probe on the pump amplitude $E_{pum}$ is striking. Therefore the considered slab in the presence of the pump and the ENZ probe can be regarded as a device allowing the pump to all-optically modulate the transmission of the probe and such all-optical control takes place even if the slab thickness is comparable with the probe carrier wavelength.

In order to prove that this all-optical modulation arises from the peculiar nonlinear wave-matter interaction in the ENZ regime, we have performed a second set of simulations for the same parameters as above except for the probe frequency, which we have set at $\omega_{pro} = 1.4 \omega_0$, so that both pulses are out of the ENZ regime. The relevant results of the simulations are reported in the right column of Fig.2. In panel (a2) of Fig.2 the well separated pump and probe spectra are reported whereas panel (b2) shows, as in the precedent situation, that the pump is linearly scattered by the slab. Panel (c2) of Fig.2 emphasizes that the spectrum of the transmitted probe pulse is actually independent on the pump amplitude $E_{pum}$, thus proving that the probe propagation through the slab is not affected by matter nonlinearity and that the slab, in the present situation, does not show any all-optical steering functionality.

The physical mechanism supporting the considered all-optical modulation can easily be grasped by exploiting the spectral separation of the two pulses to identify their own contributions to the polarization field. Specifically, in the considered pump and probe setup, both the electric and polarization fields ${\bf A} = 2 \mathop{\rm Re} \hat{\bf A}$ (where $\bf A = E,P$) have analytic signal that can be written as $\hat{\bf A} = \hat{\bf A}_{pum} + \hat{\bf A}_{pro}$ where $\hat{\bf A}_j = \exp \left(-i \omega_j t \right)\bar{\bf A}_j$  and $\bar{\bf A}_j$ is slowly varying ($j=pum,pro$). Exploiting the conditions $|{\bf P}| \ll P_s$ and $|\hat{\bf P}_{pro}| \ll |\hat{\bf P}_{pum}|$, Eq.(\ref{polariz}) yields
\begin{widetext}
\begin{eqnarray} \label{pump-pro}
\frac{{\partial ^2 {\bf{\hat P}}_{pum} }}{{\partial t^2 }} + \delta _e \omega _e \frac{{\partial {\bf{\hat P}}_{pum} }}{{\partial t}} + \omega _e^2 \left\{ {1 - \frac{3}{{P_s^2 }}\left[ {\left| {{\bf{\hat P}}_{pum} } \right|^2  + \frac{1}{2}{\bf{\hat P}}_{pum}^* {\bf{\hat P}}_{pum} ^T } \right]} \right\}{\bf{\hat P}}_{pum} &=& \varepsilon _0 \left( {\varepsilon _s  - 1} \right)\omega _e^2 {\bf{\hat E}}_{pum}, \nonumber \\
\frac{{\partial ^2 {\bf{\hat P}}_{pro} }}{{\partial t^2 }} + \delta _e \omega _e \frac{{\partial {\bf{\hat P}}_{pro} }}{{\partial t}} + \omega _e^2 \left\{ {1 - \frac{3}{{P_s^2 }}\left[ {\left| {{\bf{\hat P}}_{pum} } \right|^2  + 2{\mathop{\rm Re}\nolimits} \left( {{\bf{\hat P}}_{pum}^* {\bf{\hat P}}_{pum} ^T } \right)} \right]} \right\}{\bf{\hat P}}_{pro}  &=& \varepsilon _0 \left( {\varepsilon _s  - 1} \right)\omega _e^2 {\bf{\hat E}}_{pro},
\end{eqnarray}
\end{widetext}
i.e. the pump does not experience the presence of the probe whereas the latter is driven by the former. Due to the above small signal condition $|\hat{\bf P}_{pum}| \ll P_s$, these equations show that pump and probe experience Lorentz dielectric responses whose resonant frequencies are slightly shifted by the pump polarization field. Evidently, both the associated dielectric profiles undergo a slight drift which is responsible for a slight change of the permittivity at the carrier frequencies of the pulses. From Figs.2(a1) and 2(a2) we note that the dielectric permittivity at the pump carrier frequency is about $0.9$ so that its nonlinear change is relatively too small to trigger nonlinear sensible effects on pump propagation through the considered thin slab. Accordingly, as shown in Figs.2(b1) and 2(b2), pump propagation is unaffected by the launched pump amplitude. Very analogous is the situation where the probe carrier frequency is $\omega_{pro} = 1.4 \omega_0$ at which the dielectric permittivity is about $0.3$ (see  Fig.2(a2)) and its change due to the pump is relatively too small to produce any significant effect as confirmed in Fig.2(c2). The situation is very different when the probe is in the ENZ regime $\omega_{pro} = \omega_0$ since the pump-induced change of the dielectric permittivity it experiences is comparable to $|\epsilon(\omega_0)|$ thus entailing a marked impact of the pump intensity on the probe transmission as reported in Fig.2(c1). In addition, neglecting the contribution of $z$-component of $\hat{\bf P}_{pum}$, from the second of Eqs.(\ref{pump-pro}) we note that the presence of the pump decreases the effective resonant frequency experienced by the probe so that the real part of the dielectric permittivity at the probe carrier frequency is effectively increased thus becoming positive. Accordingly, the higher the pump intensity the more the slab departs from the metallic behavior and consequently the higher the probe transmission (see Fig.2(c1)).

It is worth estimating, in realistic situations, the peak intensity of the pump required to trigger the considered all-optical modulation of the probe. For $|{\bf P}| \ll P_s$, we have $\left( 1 + \left| \textbf{P} \right|^2/P_s^2 \right)^{-3/2} \simeq 1 - 3|{\bf P}|^2 /(2P_s^2)$ so that Eq.(\ref{polariz}) reproduces the standard anharmonic oscillator model describing the Kerr nonlinearity. Therefore, using the well-known perturbative technique (see Ref.\cite{Boyddd}) for extracting the third order nonlinear susceptibility $\chi^{(3)}$, we obtain $P_s = \epsilon_0 \sqrt{3(\epsilon_s-1)^3/ (2 \chi^{(3)})}$. From Fig.2(c1) and other simulations not reported here, we deduce that all-optical modulation occurs for pump amplitudes greater than $E_{pum} = 0.1 E_s \equiv 0.1 P_s / \epsilon_0$ whose corresponding pump peak intensity $I_{pum} = c \epsilon_0 |E_{pum}|^2 /2$ is given by $I_{pum} = 0.03 c \epsilon_0 (\epsilon_s-1)^3 /(4\chi^{(3)})$. For the above used value of $\epsilon_s$, and for the realistic nonlinear susceptibilities $\chi^{(3)} = 10^{-20}$, $10^{-19}$ and $10^{-18} m^2/V^2$, we thus obtain $I_{pum} = 1.59$, $0.15$ and $0.01 GW/cm^2$ which can be easily achievable with picosecond laser pulses.

In conclusion we have shown that all-optical modulation of a pulse can be achieved in a very thin slab by resorting to the ENZ regime. The same slab is not able to support noticeable nonlinear interaction between the pulses if they are both out of the ENZ regime. Physically, all-optical modulation is due to the oversensitive character of the ENZ regime experienced by the probe since the absolute slight change of the probe permittivity produced by the pump has a relative large impact around the ENZ frequency. The mechanism is so efficient to yield remarkable and ultrafast nonlinear pulse interaction even in a wavelength-sized slab and for relatively low optical intensities.

\section{Acknowledgements}

A.C. and C.R. acknowledge support from U.S. Army International Technology Center Atlantic for financial support (Grant No. W911NF-14-1-0315).


\end{document}